\documentclass[runningheads]{llncs}
\usepackage[T1]{fontenc}
\usepackage{graphicx}
\usepackage{color}
\usepackage{colortbl, booktabs}
\usepackage{xcolor}

\usepackage{comment}

\usepackage{changepage} % for the adjustwidth environment
\usepackage{amsmath} % for the math equations

\usepackage{multirow}
\usepackage{tabularx}
\usepackage{makecell}
\usepackage{hhline}
\usepackage{diagbox}
\usepackage{xcolor}

\usepackage{float}
\floatstyle{plaintop}
\restylefloat{table}

\usepackage{url}

\newif\ifdraft
 \drafttrue
 \draftfalse
\ifdraft
  \newcommand{\todo}[1]{\textsf{\textbf{\textcolor{red}{[TODO: #1]}}}}
  \newcommand{\key}[1]{\textsf{{\textit{\textcolor{red}{#1:}}}}}
  \newcommand{\takuya}[1]{\textsf{\textcolor{blue}{\textbf{Takuya:} \textit{#1}}}}
  \newcommand{\yuri}[1]{\textsf{\textcolor{orange}{\textbf{Yuri:} \textit{#1}}}}
\else
  \newcommand{\todo}[1]{}
  \newcommand{\key}[1]{}
  \newcommand{\yuri}[1]{}
  \newcommand{\takuya}[1]{}
\fi
\begin{document}
\title{Towards Multi-Stakeholder Evaluation of ML Models: A Crowdsourcing Study on Metric Preferences in Job-matching System}
\titlerunning{Towards Multi-Stakeholder Evaluation of ML Models}
\author{Takuya Yokota\orcidID{0009-0000-0470-5313} \and
Yuri Nakao\orcidID{0000-0002-6813-9952}}
\authorrunning{T. Yokota and Y. Nakao}
\institute{Fujitsu Limited, Kawasaki-city, Japan \\
\email{\{yokota-takuya, nakao.yuri\}@fujitsu.com}}

\maketitle              % typeset the header of the contribution
\begin{abstract}
%% The abstract should briefly summarize the contents of the paper in 150--250 words.
While machine learning (ML) technology affects diverse stakeholders, there is no one-size-fits-all metric to evaluate the quality of outputs, including performance and fairness. Using predetermined metrics without soliciting stakeholder opinions is problematic because it leads to an unfair disregard for stakeholders in the ML pipeline. In this study, to establish practical ways to incorporate diverse stakeholder opinions into the selection of metrics for ML, we investigate participants' preferences for different metrics by using crowdsourcing. We ask 837 participants to choose a better model from two hypothetical ML models in a hypothetical job-matching system twenty times and calculate their utility values for seven metrics. To examine the participants' feedback in detail, we divide them into five clusters based on their utility values and analyze the tendencies of each cluster, including their preferences for metrics and common attributes.  Based on the results, we discuss the points that should be considered when selecting appropriate metrics and evaluating ML models with multiple stakeholders.

% \keywords{First keyword  \and Second keyword \and Another keyword.}
\end{abstract}
\section{Introduction}

% 機械学習はいろいろな社会的な決定に使われています。
% 機械学習の出力は多くの人に影響を与える。
% 社会的決定には多くのステークホルダーが含まれている。
% 機械学習システムの設計はこれらのステークホルダーを交えて考えるべきである。
% \todo{cite appropriate references in each part that need them}

% \todo{decide which to use, convert, aggregate, integrate}

% \todo{add the requirement for multi-stakeholder evaluation of ML models and how this work connects to the multi-stakeholder evaluation > yuri}

As machine learning (ML) technologies have impacts on a diverse range of stakeholders,
% ~\cite{Miller2022,nakao2023TowardsResponsible, KIM_2023_DoStakeholder, Yokota2022,Nakao2023Stakeholder,Delecraz2022}, 
there have been ongoing efforts to improve the transparency of ML systems by including a variety of stakeholders in the evaluation of ML models~\cite{Miller2022,nakao2023TowardsResponsible,nakao2022toward,KIM_2023_DoStakeholder,Delecraz2022}.
Because the evaluation of ML models requires a variety of metrics, including accuracy or fairness~\cite{Madaio2020CoDesigning}, some studies have aimed to incorporate stakeholders in the process of selecting metrics and deciding how much weight to give to them~\cite{Yokota2022,nakao2023TowardsResponsible,Ahn2020FairSight}.
For example, one approach is to have stakeholders directly adjust the parameters of the model themselves~\cite{nakao2022toward}, while another is to survey their preferences for each metric~\cite{Yokota2022}.

% Such efforts have been made to avoid an ML model being defined by the administrators in a top-down way, which leads to ignoring the diverse stakeholders' opinions.
% which leads to an obstacle to the practical use of ML.

% , including non-experts~\todo{cite} as well as AI experts~\todo{cite}, 

On the other hand, it is difficult to fairly aggregate stakeholders' perspectives on the ML model evaluation. 
While some studies pre-defined the stakeholder groups to collect feedback from the stakeholder groups' points of view~\cite{nakao2023TowardsResponsible,Cheng2021Soliciting}, different opinions exist within the same stakeholder group~\cite{nakao2023TowardsResponsible,Nakao2022TowardInvolving}, and there has been a concern that even in a stakeholder group, a minority opinion in the group can be ignored~\cite{Green_2001_Community,Greenwood_2009_Iused}.
Then, 
How should stakeholder feedback on ML models be collected? How should stakeholder groups be determined?

To address these questions, as a first step toward evaluating ML models including diverse stakeholders, we propose a method for examining what metrics people tend to prefer, how these tendencies can be grouped, and what attributes people are likely to include in each group.
In this study, we used crowdsourcing to assign 837 participants residing in the U.S. the task of showing two outputs from hypothetical ML models for job matching and choosing the model they thought was better. Based on the results of this task, we determined people's preferred metrics and clustered them based on their preferences. Furthermore, we analyzed what attributes people tended to include in each cluster, i.e., stakeholder groups.
In this study, we address the following two-fold research questions:

\begin{description}
    \item[RQ1] How are participants' preferences for ML performance metrics clustered?
    % A questionnaire-based method can extract potential stakeholders' preferences? \yuri{}
    \item[RQ2] What demographic attributes have a significant effect on the preferences clustered with our method?
    % A questionnaire-based method can extract potential stakeholders' demographic attributes in order to explain their preferences? \yuri{What demographic attributes have significant effects on the clusters of the preferences with our method?}
\end{description}

The contributions of our study are as follows:
\begin{itemize}
    \item A new method to analyze the characteristics of stakeholders who evaluate ML models in similar ways.
    \item The analysis of what tendencies in attributes people who make a particular evaluation of ML models share in the domain of job matching.
    \item Implication toward fair multi-stakeholder evaluation of ML models based on the results.
\end{itemize}

\section{Related Work}
% \yuri{\#1. O2. The motivation of the paper could be more strongly grounded in prior literature ? what does it mean for stakeholders to participate in the development or performance evaluation of ML models?}
% In this section, we characterize the research domain to which this paper contributes and introduce related studies. 

\subsection{Evaluation Metrics for Machine Learning Models}
% Fair Machine-learning Practice
\label{ssec:2_fair_ml}

% メトリクスは多様。
Existing studies have emphasized that evaluating ML models involves carefully choosing metrics, such as performance and fairness, tailored to specific situations. While general metrics like Accuracy, Precision, and Recall are broadly applicable across a range of domains~\cite{powers2020evaluation}, Sokolova and Lapalme~\cite{Sokolova2009Systematic} discussed the relevance of these metrics can vary depending on the application.
Additionally, beyond performance measures, fairness metrics, such as Disparate Impact~\cite{chouldechova2017}, are gaining importance. 
% in such as recidivism prediction instruments. 
ML model operators need to carefully choose which of these various metrics to use based on their purpose and ethical considerations.
% This diversity in metrics underscores the need to carefully choose those that align with the intended purpose and ethical considerations of the specific machine learning application.

% 人々は異なる優先度をつける。
% \subsection{Human Perception of Machine-learning Models} \label{ssec:2_human_perception}
Moreover, people's priorities for these metrics differ based on the context and stakes of decision-making~\cite{Holstein2019,Srivastava2019,Berkel2021,Harrison2020}.
% People's perception of metrics changes depending on the decision-making situations.
For example, for high-stake decision-making situations, e.g., cancer-detection predictions, people prioritize accuracy over fairness compared with a flu-symptom-severity prediction (low-stake situations)~\cite{Srivastava2019}.
Harrison et al. found that people see a model as biased when it has disparate false-positive rates that disadvantage African American defendants in a recidivism risk-assessment system~\cite{Harrison2020}. 
Furthermore, the perception of metrics changes on the basis of people's education level~\cite{Wang2020} and characteristics of the tasks~\cite{Harrison2020}. 
Previous studies quantified the change or difference in participants' attitudes toward ML systems~\cite{Srivastava2019,Berkel2021}. 
They used explanations about the mechanisms of algorithms or performance metrics that express the behavior of an ML system as proxies.

In this study, we selected Accuracy, Specificity, Sensitivity (Recall), and Precision as evaluation metrics to understand stakeholder preferences in the context of a job-matching system.
Additionally, as fairness metrics, we pick up Disparate Impact (DI)~\cite{chouldechova2017}, Equalized Odds (EO)~\cite{Hardt2016}, and Counterfactual Fairness (CF)~\cite{kusner2017} as evaluation metrics to collect the preferences relate to fairness as well. This allows for a comprehensive exploration of how stakeholders evaluate the performance and fairness of the job-matching system.
% by assessing various aspects of the system.
Additionally, this study extends the existing studies about people's perception of the metrics by including the viewpoint that they are stakeholders who are potentially affected by the system rather than just the general public. This approach is selected to gather diverse insights on modeling user preferences for metrics.

\subsection{Stakeholder Models in Multi-Stakeholder Machine Learning Systems} 
% \subsection{Stakeholders of machine learning based Decision Support Systems\todo{reconsider}}
\label{ssec:2_stakeholder_models}
% ステークホルダーに関する研究：機械学習の関連研究、ジョブマッチングの関連研究。

\todo{modify based on the revised introduction.}

\yuri{if possible, add this content $\rightarrow$ Especially in the context of job-matching systems, \todo{extend}}

% The integration of ML into societal applications has necessitated understanding the diverse stakeholders involved. 
% Narayanan~\cite{Narayanan2018} emphasized the need for understanding stakeholder-specific preferences in social decision support systems.

Not only the human factors in the choice of ML metrics but the importance of considering a diverse range of stakeholders in the development and implementation of business applications has been emphasized.
Mitchell et al.~\cite{Mitchell_2019_ModelCards} and Lee et al.~\cite{Lee_2019_WeBuildAI} emphasized the importance of acknowledging the stakeholders of ML systems. 
In business applications, the identification of key stakeholders is crucial. Preece et al.~\cite{Preece2018} mention including decision-makers, end users, and those impacted by decision-support systems. Golbin et al.~\cite{Golbin2019} specify including data scientists, developers, business sponsors, and regulators as pivotal stakeholders. 

\todo{add the first paragraph}
% \todo{Elaborate our approach compared to recommender systems}
There have also been existing studies to include stakeholders' viewpoints in the operation and development of machine learning and other artificial intelligence systems.
Burke et al.~\cite{Burke2016} designed a method to incorporate stakeholders' preferences into ML models for recommender systems, emphasizing the need to balance benefits among predefined stakeholders like item providers, receivers, and system owners. Similarly, Zheng and Toribio~\cite{Zheng2019Personalized} advocated for balancing multiple stakeholders' needs in recommender systems, proposing a model that enhances transparency by explaining key parameters affecting stakeholders' utility. In a recidivism prediction system, Narayanan~\cite{Narayanan2018} highlighted how various stakeholders of a recidivism prediction system prioritize different performance metrics. Extending this approach, Yokota and Nakao~\cite{Yokota2022} employed a utility-based method to extract crowdsourcing workers' preferences for ML model evaluation metrics in a similar system.

% Identifying key stakeholders, such as decision-makers, end users, and those affected by decision-support systems, is vital in business applications, as noted by Preece et al.\cite{Preece2018} and Golbin et al.\cite{Golbin2019}.

% ステークホルダーはシステムごとに考えられる必要がある。
However, stakeholder identification should be able to be tailored to each system so that we can appropriately consider all relevant stakeholders in the development and implementation process, not only by predetermining the stakeholder groups.
In our crowdsourcing study, we analyze how the stakeholders' demographic attributes influence their preferences in ML systems and clarify what tendencies in attributes of people who have a particular tendency in ML model evaluation. 
Through this study, we aim to clarify the need to include the demographic and domain-specific attributes of the stakeholders to understand their needs and preferences in more detail.

% In our crowdsourcing study, 
% we focus on a job-matching system where many participants are likely to belong to either the employee or employer category, 
% thereby covering a broad range of potential stakeholders.
% This approach allows us to extract opinions with realistic stakeholder interests, 
% even in hypothetical scenarios like those in our study.

% However, the existing literature overlooks how the stakeholders' demographic attributes and beliefs influence their preferences in ML systems, indicating a need to include these factors for a more detailed understanding of their needs and preferences.

% To address this gap, 
% our study incorporates the stakeholders' demographic attributes and beliefs into stakeholder models. 
% This approach seeks a more comprehensive analysis of stakeholder preferences.
% Additionally, because our study focuses on stakeholders of a job-matching system, primarily non-expert users, demographic factors like information technology (IT) education can influence stakeholders' preferences in ML systems. 
% We aim to mitigate the impact of knowledge disparity by designing a baseline explanation of a job-matching system tailored for individuals without IT-related education. 

\yuri{Baba et al. 2020 commented out}

\section{Method}

Our research is a crowdsourcing study to explore which demographic attributes of participants affect their preferences for machine learning models. We showed participants a hypothetical job matching scenario and asked them to choose the desired model output through a set of discrete choice tasks. This way, we extracted their preferences as a balance of seven performance and fairness metrics using utility funtion. Based on these seven metrics, we clustered their preference patterns and identified the characteristics of each cluster. We then 
analyzed the relationship between the participants' attributes and their preferences for the machine learning models.

\subsection{Procedure}

% \takuya{9. Review the explanation to distinguish what is told to the participants, and what procedure is actually done: DONE}
% We conducted our crowdsourcing research with the following procedure via Amazon Mechanical Turk (MTurk). 
% After agreeing on the consent form, participants were shown the scenario. 
% In the scenario, we described the current state of AI use in job matching, a hypothetical use case, as follows:
We conducted our crowdsourcing research using Amazon Mechanical Turk (MTurk) with the following procedure. 
After consenting to the form, participants were shown the scenario. 
In the scenario, we described the current state of AI use in job matching, presenting a hypothetical use case, as follows:
\begin{quote}
    In the United States, many employers and job seekers use job-matching websites. Currently, Artificial Intelligence (AI) systems support employers’ decisions by showing an optimized list of candidates. An AI system uses applicants’ resumes and CVs as training data to calculate its AI model to predict the possibility of matching. The training data for the AI model includes several attributes, such as applicants’ education, job history, and skills. The AI system calculates the applicants’ expertise based on the training data and predicts the likelihood of successful matching as “Likely to be hired” or “Not likely to be hired.”
\end{quote}

After explaining the scenario, we included an attention check task to ensure participants paid appropriate attention to the explanations. 
Participants were then asked to select one AI model from two options and repeated this process twenty times to accurately capture their preferences.
During each selection task, participants were shown the ten applicants' race, expertise, actual matching results, and prediction results from two ML models. 
We note that we used hypothetical ML model results, generated randomly, in the selection task instead of actual models trained with data. We used these hypothetical results to efficiently elicit participants' preferences.

We elaborate on this main task in~\ref{ssec:3_MainTask}. After completing the main task, we collected demographic information from the participants. This demographic data was used to investigate the relationship between stakeholders' attributes and their preferences, as detailed in~\ref{ssec:3_Association_Analysis}.

\subsection{Participants} \label{sssec:3_participants}
The participants are recruited through MTurk. 
We limit participants who live in the U.S. 
The participants have to complete at least 500 `human intelligence tasks' (HITs) and have at least a 95\% HIT approval rate to guarantee the reliability of their answers. 
Our study was conducted from 20th to 23rd December 2022. 
% \takuya{10. Justify the amount of compensation: DONE}
We compensated the participants US\$2.42 for their time. This amount of compensation is calculated based on the U.S. minimum wage, which is US\$7.25 per hour, which is also the case in the related study~\cite{Lee2018}.
The average time they needed to complete the survey was 18.2 minutes (\textit{standard deviation (SD)} $=34.50$). 
$1100$ people responded to the survey. 
We omitted $263$ participants in total, including those who did not finish the survey ($N=14$), whose reCAPTCHA scores\footnote{reCAPTCHA is an optional function of Qualtrics. See the documents on reCAPTCHA Enterprise for details about the reCAPTCHA score. \url{https://cloud.google.com/recaptcha-enterprise}} were under $0.5$, which indicates the responses might have been made by bots ($N=22$), who gave contradictory answers to multiple choice questions ($N=38$), and who failed the attention-check questions ($N=189$). 
As a result, $837$ participants were left.
% 

% \todo{the following part in this subsubsection is too long. Summarize if needed.}
% \takuya{4. Justify why the attributes are chosen: DONE}
Next, we show the demographic information of participants as follows. The largest age group was 25--34 years old (62.1\%), and the second largest was 35--44 years old (20.3\%). 
For other age groups, 18--24 was 3.3\%, 45--54 was 8.2\%, 55--64 was 5.4\%, 65+ was 0.6\%. 
Those identifying themselves as female were 33.2\%, and as male were 66.8\%. 
No one identified themselves as `gender non-binary' or `others not listed.' 
For the ethnic group, 89.0\% were White, 3.1\% were African American, 2.0\% were Asian, 1.7\% were Hispanic, 4.2\% were Native American or Alaska Native, and there were no Native Hawaiians or Pacific Islanders. 
Regarding income, 6.3\% earned `less than \$25,000,' 28.2\% earned `\$25,000 to \$49,999,' 32.7\% earned `\$50,000 to \$74,999,' 25.2\% earned `\$75,000 to \$99,999,' 6.0\% earned `\$100,000 to \$149,999,' and 1.6\% earned `\$150,000 or more.'
The demographic attributes obtained are adapted from the questionnaire used in the previous studies~\cite{Lee2018,Lee2019,Yokota2022}.

Additionally, we asked the information on the participants possibly related to our scenario.
First, for the participants' work situations, all the participants were employed: 96.3\% had a full-time job, 2.9\% had a part-time job, and 0.8\% were students with full-time jobs. 
Their occupations were dominated by a few categories: 36.9\% were professional, 34.2\% were managerial, and 18.8\% were self-employed. 
Regarding education level, most had a `4-year degree' (66.2\%), and the second largest group had a `professional degree' (14.2\%). 
Other groups were `high school graduates' (12.1\%), `less than high school ' (2.2\%), `some college' (3.8\%), and `2-year degree' (1.6\%). 
Thus, 82.0\% of the participants were above `2-year degree,' which means they were well educated. 
% \takuya{13. Remove the descriptions about IT-related degree and ML experience: DONE}
Regarding political affiliation, 39.5\% of the participants were Republicans, 49.0\% were Democrats, and 11.4\% were Independent.

    \begin{figure*}[h]
        \centering
        \includegraphics[width=0.7\linewidth]{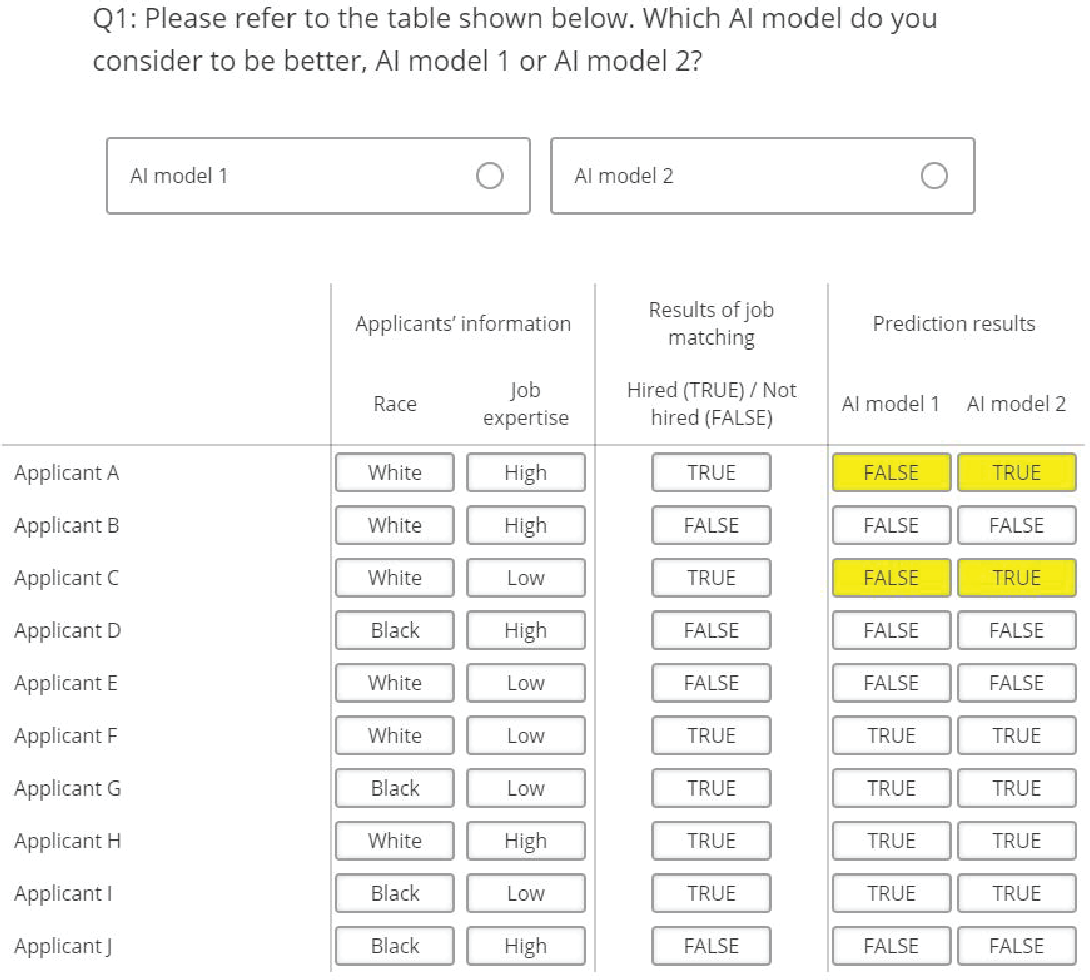}
        \caption{Discrete choice task: ten job applicants' information, race as Black or White, expertise as high or low, and actual results of matching as hired (true) or not hired (false), with the prediction results of two hypothetical ML models. Only two rows had different sets (with a yellow background for emphasis) to reduce cognitive overload. Both rows and values were randomized in each task.}
        \label{fig:3_method_task_desc}
    \end{figure*}

% \subsection{Questionnaire Survey Design} 

\subsection{Main Task} \label{ssec:3_MainTask}

For the main task, we showed participants information about ten job applicants along with the prediction results of two hypothetical ML models. We then asked them to choose the better of the two models. Participants made this choice twenty times, each time with a different combination of ML models. This allowed us to calculate each participant's preference using a utility function based on the discrete choice model while also avoiding cognitive overload.
% For the main task, we showed participants ten job applicants' information with the prediction results of two hypothetical ML models and asked them to choose the better of the two models. 
% We have the participants choose their preferred model twenty times with a different combination of ML models to calculate each participant's preference with a utility function using the discrete choice model as well as avoiding cognitive overload. 
% By having the participants choose their preferred model 20 times with a different combination of ML models, we can calculate each participant's preference with a utility function using the discrete choice model. 
% Twenty times was determined to balance avoiding cognitive overload and extracting enough information to calculate the preference. 
Figure \ref{fig:3_method_task_desc} shows a question presented to the participants in this task. 
In the question, applicants' race, job expertise, results of job matching, and prediction results from hypothetical ML models 1 and 2 are shown. 
% For this question, five items were shown for each job applicant, i.e., race, job expertise, results of job matching, and prediction results from ML models 1 and 2. Race and job expertise were the applicants' information. 
% The ``results of job matching'' show if the applicant was actually hired or not, and the ``prediction results'' are the results from different ML models. 
% The results of job matching were the true results of whether the applicant was actually hired or not. 
% The prediction results are the results from different ML models. 
For ``race,'' there were two values, i.e., Black and White. For ``job expertise,'' there were also two values, i.e., ``high'', which indicates the applicant was qualified for the job in terms of job experience and certifications, and ``low'', which means the applicant was less qualified for the job. 
For ``results of job matching,'' there were two values, i.e., TRUE and FALSE, which are based on the actual results of past job matching. 
% TRUE means the applicant was hired from job matching, and FALSE means s/he was not hired. 
TRUE means the applicant was hired, and FALSE means they were not hired.
% Finally, the ``Prediction results,'' are shown from the two ML models, i.e., ML models 1 and 2, that predict the results of job matching using applicants' race, education, job experience, job history, and skills. 
% The results are shown with two values; TRUE and FALSE. 
% TRUE is shown when the applicant is predicted likely to be hired, and FALSE is shown if s/he is not likely to be hired. 
Finally, the ``Prediction results'' from the two ML models (ML model 1 and ML model 2) were shown. These models predict the job matching results using applicants' race, education, job experience, job history, and skills. The prediction results were shown as TRUE (likely to be hired) or FALSE (not likely to be hired).

% \takuya{8. What is `randomisation' of labels? Completely random or some swapping? It would be good to give a bit more detail here for replicability: DONE}

% In our setting, the two hypothetical ML models are supposed to be trained differently and result in a different set of prediction results.
% To make the results from the hypothetical models, we randomly select two rows out of ten rows, i.e., applicants, and the prediction results in these selected rows are made to differ between ``AI model 1'' and ``AI model 2.'' We determine if these values are true or false randomly. The remaining eight rows are populated with values that are consistent with the column ``Results of job matching.''
In our setting, participants were asked to compare two AI models and choose the one with better results. However, these two models do not actually exist. Instead, we created the questions by randomly selecting two out of ten applicants and making their prediction results differ between ``AI model 1'' and ``AI model 2.'' The true or false values for these predictions were determined randomly. The remaining eight rows were consistent with the ``Results of job matching'' column. This setup allowed us to observe participants' reactions to the differing results and calculate their preferences using a utility function.

\subsection{Utility Function} \label{ssec:3_utility_function}
\subsubsection{Performance Metrics}
On the basis of the responses to the questionnaire, we calculated the participants' preferences as the set of preference values for ML metrics. The metrics are categorized into performance metrics and fairness metrics. As performance metrics, we use accuracy, specificity (or true-negative rates), sensitivity (or recall), and precision. 
As fairness metrics, we used disparate impact (DI), equality of opportunity (EO), and counterfactual fairness (CF). 
We give details of these seven metrics as follows:
% \takuya{13. For the seven metrics, why not define formally?: DONE}
    \begin{description}
        \item [Accuracy]: Accuracy is calculated as the sum of the applicants who were correctly predicted divided by the total number of applicants. There are two types of correct prediction; where an applicant was predicted as `TRUE (likely to be hired)', i.e., true positive, and where an applicant was predicted as `FALSE (NOT likely to be hired)', i.e., true negative. 
        \item [Specificity (or True Negative Rates)]: Specificity is calculated as the ratio of the number of applicants correctly predicted as FALSE to the total number of applicants who actually were NOT hired (FALSE).
        \item [Sensitivity (or Recall)]: Sensitivity is calculated as the ratio of the number of applicants correctly predicted as TRUE to the total number of applicants who actually were hired (TRUE).
        \item [Precision]: Precision is calculated as the ratio of the number of applicants correctly predicted as TRUE to the total number of applicants who were predicted as TRUE.
        \item [DI]: DI is a group fairness metric and is calculated as the ratio of the proportion of Black applicants (regarded as an underprivileged group) who obtained a favorable outcome (TRUE prediction) to the proportion of White applicants (regarded as a privileged group) who obtained a favorable outcome~\cite{chouldechova2017}. When we denote the protected attributes as $r$ (where $r=b$ for Black and $r=w$ for White), DI is as follows:
        \begin{equation}
            DI = \frac{P(\hat{y}=1 | r=b)}{P(\hat{y}=1 | r=w)}
        \end{equation}
        % DI is satisfied with respect to the protected attribute $r$ (where $r=b$ for Black and $r=w$ for White) if the condition below is met:
        % \begin{equation}
        %     P(\hat{y}=1 | r=b) = P(\hat{y}=1 | r=w)
        % \end{equation}
        Here, $\hat{y}=1$ denotes the predicted label indicating a favorable outcome.
        When $DI = 1$, the model is thought of as completely fair with respect to the protected attribute.
        \item [EO]: EO is a group fairness metric and is calculated as the ratio of the probability of Black applicants (underprivileged group) correctly predicted as TRUE in the total number of Black applicants who actually were hired to the probability of White applicants (privileged group)  of correctly predicted as TRUE in the total number of White applicants who actually were hired~\cite{Hardt2016}. 
        When we denote the protected attributes as $r$, EO is as follows:
        \begin{equation}
            EO = \frac{P(\hat{y}=1 | r=b , y=1)}{P(\hat{y}=1 | r=w , y=1)}
        \end{equation} 
        % EO is satisfied with respect to the protected attribute $r$ if the following condition holds:
        % \begin{equation}
        %     P(\hat{y}=1 | r=b , y=1) = P(\hat{y}=1 | r=w , y=1)
        % \end{equation} 
        Here, $y=1$ represents the actual label if the applicant is hired (TRUE). When $EO = 1$, the model is thought of as completely fair with respect to the protected attribute.
        \item [CF]: CF is an individual fairness metric. In our study, CF is defined as the ratio of applicants whose job expertise is high and predicted as TRUE to the total number of applicants. This is because, in the concept of individual fairness, an ML model is fair if it gives similar predictions to similar individuals~\cite{kusner2017}. The applicants should be determined only by their job expertise, not by their demographic attributes. Therefore, if ``Job expertise''=``High``, then the prediction of the model $\hat{y}$ should be 1.
    \end{description}

% ~\cite{Zafar2017}

% \takuya{modified $\rightarrow$}
\noindent Here, we pick group fairness metrics, i.e., DI and EO, because it is possible that there are differences in the awareness of employees' interests between direct recipients of the decisions, i.e., employees and employers. For example, while employees might prioritize these fairness metrics, employers do not prioritize them to the same extent. Additionally, the individual fairness metric, CF, is chosen because it is assumed that highly skilled participants might prioritize occupational suitability and believe that fairness should be determined solely on the basis of job competence. 
% Unlike DI and EO, CF does not emphasize collective fairness but individual job suitability.
% \yuri{need to reason for the conventional fairness metrics should be added?}

% \takuya{\#13 O2: The procedures of how the analyses were conducted could be further clarified. Currently, many aspects of the analyses were unexplained. For example, it is unclear why the metrics in 3.4.2 were calculated and how they were used in the analyses. }

Additionally, please note that there are trade-offs between some metrics while others overlap. For instance, specificity and sensitivity (recall) are in trade-off relations~\cite{Rodolfa2021,Hardt2016}. Recall (sensitivity) and precision are also in trade-off relations. However, accuracy may increase when specificity (true negative rates), sensitivity (recall), or precision increases. We used the essential metrics for both performance and fairness to determine a participant's nuanced preferences.

\subsubsection{Discrete Choice Model} 
To calculate the participants' preferences, we used the discrete choice model~\cite{kjaer2005,mcfadden1974conditional} to calculate the preferences to maximize the utility for each participant on the basis of their choices of models. 
When participant $i$ chooses model $j$, the utility function $U_{ij}$ should satisfy the following inequality:

\begin{equation}
U_{ij} > U_{ik} \;\; \forall j \neq k \end{equation}

\noindent 
% Let \begin{math}U_{ij}\end{math} be the utility function of the model which participant $i$ selects choice $j$, 
In the discrete choice model $U_{ij}$ is expressed as follows:

    \begin{equation}
      U_{ij} = V_{ij} + \epsilon_{ij}
    \end{equation}

\noindent where \begin{math}V_{ij}\end{math} is the deterministic (observed) part and \begin{math}\epsilon_{ij}\end{math} is the random (unobserved) part. 
In this study, $V_{ij}$ was a function of observed variables $x_{ij}$ (accuracy as $Acc$, specificity as $Spe$, sensitivity as $Sen$, precision as $Pre$, and $DI$, $EO$, $CF$):

    \begin{equation}
      \begin{split}
      V_{ij} &= \alpha + \beta_{Acc_i} x_{Acc_{ij}} + \beta_{Spe_i} x_{Spe_{ij}} + \beta_{Sen_i} x_{Sen_{ij}} \\
          &\quad
          + \beta_{Pre_i} x_{Pre_{ij}}  + \beta_{DI_i} x_{DI_{ij}} + \beta_{EO_i} x_{EO_{ij}} + \beta_{CF_i} x_{CF_{ij}} 
      \end{split}
    \end{equation}

\noindent The parameters $\beta$ are the preference values of the seven metrics and the same for all models within each individual, and \begin{math}x_{ij}\end{math} are the metric values of the $j$th model that participant $i$ selects. 
For example, when there are only two models, ML model 1 and ML model 2, the likelihood of participant $i$ choosing ML model 1, \begin{math}L_{i1}\end{math}, is as follows: 

    \begin{equation}
      L_{i1} = Prob.[ U_{i1}] = \displaystyle \frac{ \exp\{U_{i1}\} }{ \sum_{k=1}^{2} \exp\{U_{ik}\} },
    \end{equation}

% \takuya{Modified->}
\noindent The log-likelihood $L_{i}$ of participant $i$ after the $l$th tasks ($l=1,\ldots,20$) is defined as the product of the likelihood of single selection $L_{ij}$, which we denote as

    \begin{equation}
        \displaystyle L_{i} = \log \prod_{l=1}^{20} L_{ij}
    \end{equation}

    % \begin{equation}
    %   \setlength{\abovedisplayskip}{3pt}
    %   \setlength{\belowdisplayskip}{3pt}
    %   \log \prod_{l=1}^{20} L_{i1} = \sum_{l=1}^{20} U_{i1} - \sum_{l=1}^{20} \log \left[ \sum_{k=1}^{2} \exp\{U_{ik}\} \right]
    % \end{equation}

\noindent Estimation of the discrete choice model is based on a maximum likelihood procedure to find the parameters of $\beta$, the preference values of the seven metrics, that maximize $L_{i}$.

% Next, by using cluster analysis, we examine how the metric prioritization of participants can be categorized.

\subsection{Data Analysis}
\subsubsection{K--means Cluster Analysis} \label{ssec:3_k_means}
To investigate the patterns in participants' preferences, we clustered the preferences with k-means clustering~\cite{hartigan1979algorithm}. 
In k-means clustering, $k$ points are first selected as initial centroids in the feature space. 
As the feature space, we set the 7-dimensional space consisting of the preferences for seven performance ML metrics, i.e., the preferences for accuracy, specificity, sensitivity, precision, DI, EO, and CF. 
The $k$ clusters are then formed by assigning each data point to its closest centroid. 
In this study, the data points were the preferences of the participants ($N=837$). 
The centroid of each cluster is next moved to the center of the points that belong to each centroid. 
Forming clusters and computing centroids are then repeated until the positions of centroids stop moving. 
Through the clusters resulting from k-means clustering, we analyzed the patterns of preferences for performance metrics. 
Using the patterns, we analyzed the distinctive demographic attributes of participants when the preferences were divided into clusters.

In k-means clustering, we initially have to set the number of clusters $k$. 
We use the elbow method to determine $k$~\cite{bholowalia2014ebk}. 
With this method, the ideal number of groups is determined as the number at which, if more groups are added, there is no significant difference in the reduction of variance within groups. 
We used the Python scikit-learn library with the default setting to apply the k-means clustering algorithm and elbow method\footnote{https://www.scikit-yb.org/en/latest/api/cluster/elbow.html}. 
With the library, we identified that $k=5$ is the ideal number of clusters. 

\subsubsection{Association Analysis} \label{ssec:3_Association_Analysis}

% \takuya{15. Need additional analysis, describe the method of the analysis: DONE}

In this study, we use association analysis to explore the relationships between participants' demographic attributes and clusters. 
Association analysis is a data mining technique that is used to discover relationships between items within a dataset~\cite{McNicholas2008,hussein2015}. 
% and is primarily employed in market basket analysis, recommendation systems, and inventory management.
Out of this technique, we used lift values as our evaluation metric. Lift values provide a measure of how strongly two items are associated. In our case, we investigate how one value in the demographic attributes is associated with a specific cluster to analyze the relationship between participants' demographic attributes and clusters. In other words, we consider a value in participants' demographic attributes and a specific cluster as two different items. 

The formula for calculating the lift value is as follows:
    \begin{equation}
        \text{Lift}(A, B) = \frac{P(A \cap B)}{P(A) P(B)}
    \end{equation}
Where $P(A \cap B)$ is the probability of both A and B occurring together, $P(A)$ is the probability of A occurring, and $P(B)$ is the probability of B occurring~\cite{McNicholas2008}. If the lift value is greater than 1, it suggests that items A and B have a positive association with each other~\cite{hussein2015}. Conversely, a lift value less than 1 indicates a negative association between the items.
By using the lift value as a measure, we can quantify the strength of the relationship between a specific value in participants' demographic attributes and a cluster, leading to a more structured analysis.

It is important to note that there is no universal benchmark that indicates a significant ``high'' or ``low'' lift value.
In our analysis, we set threshold values at 0.750 for low lift value and 1.250 for high lift value for interpreting lift values, aiming to identify strong and weak associations between demographic attributes and clusters.
These thresholds were chosen to capture attributes that deviate by at least 25\% from a lift value of 1, which would indicate no association. 
% The inverse of 0.750 is 1.333, allowing us to examine both ends of the association spectrum symmetrically.
% By setting these thresholds, we aim to provide a more nuanced understanding of the relationships between demographic attributes and clusters. 
This approach allows for a comprehensive interpretation of the data, particularly when considering the influence of smaller cluster sizes.
% the potential for skewed distributions and 
In this study, a high lift value indicates that a particular attribute increases the likelihood of being in that cluster by a factor corresponding to the lift value. For example, a lift value of 2 would mean that the attribute doubles the probability of a participant belonging to that cluster.

\section{Results}

% https://library.sacredheart.edu/c.php?g=29803&p=185931
% An introductory context for understanding the results by restating the research problem that underpins the purpose of your study.
% This paper aims to show that a questionnaire-based method can extract stakeholders' preferences for AI models.
% Additionally, in the course of requirements extraction, we identify stakeholders by their demographic attributes in order to clarify ``Who benefits from what?'' which provides a mathematical explanation of the stakeholders.
% Therefore, we addressed the following research questions:
% % \vspace{-1mm}
% \begin{description}
%     % \setlength\itemsep{0em}
%     \item[RQ1] Can a questionnaire-based method can extract potential stakeholders' preferences?
%     \item[RQ2] Can a questionnaire-based method can extract potential stakeholders' demographic attributes in order to explain their preferences?
% \end{description}

% A summary of your key findings arranged in a logical sequence that generally follows your methodology section.
We first summarize the clustering results and then examine the demographic differences between clusters using the association analysis. 

% Inclusion of non-textual elements, such as, figures, charts, photos, maps, tables, etc. to further illustrate the findings, if appropriate.
% OK

    \begin{figure*}[h]
        \centering
        \includegraphics[width=\linewidth]{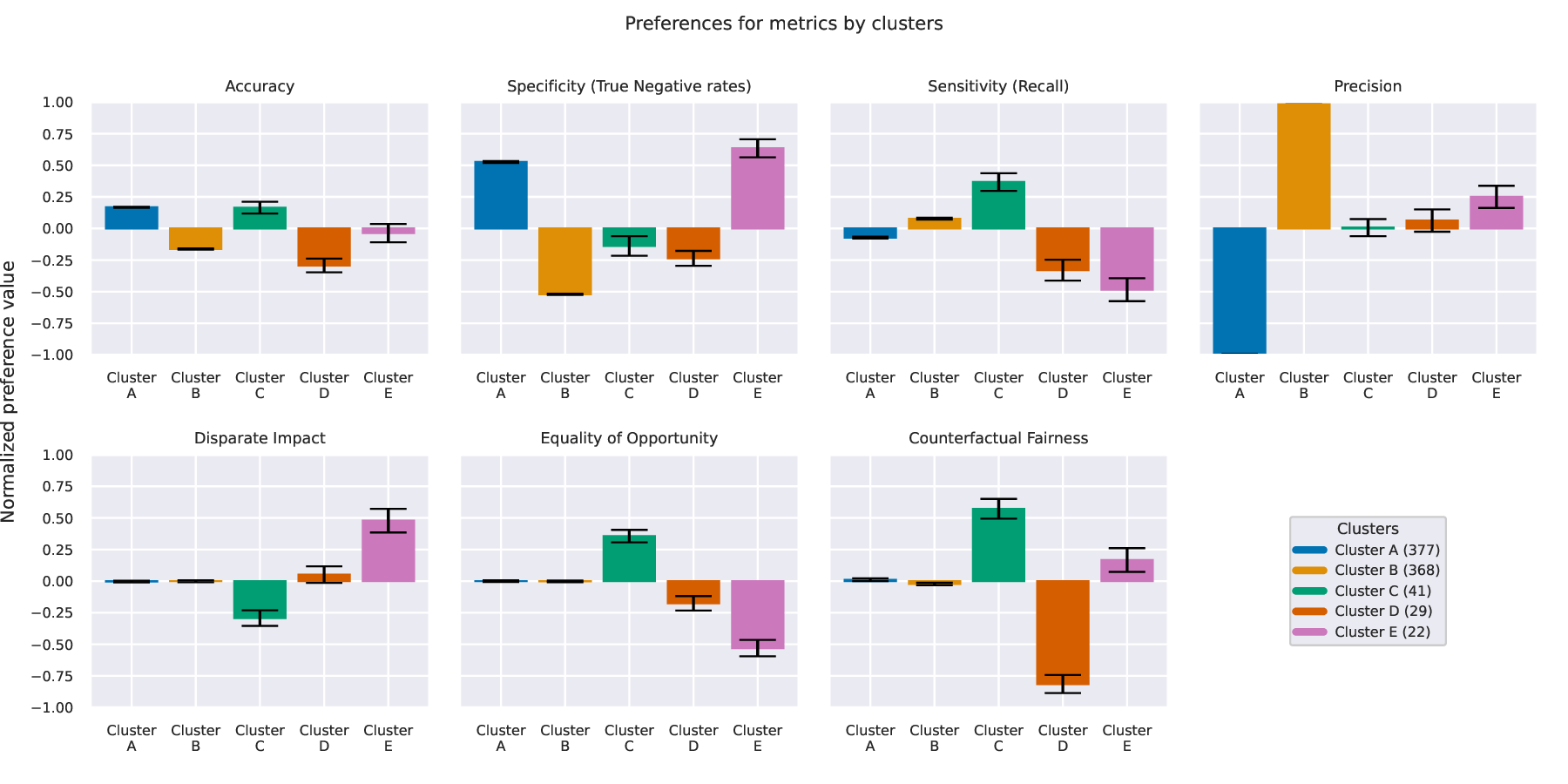}
        \caption{Result of k-means clustering. Each color represents the average preference value of participants assigned to each cluster. The vertical axis represents preference values that have been normalized to a range of $-1$ to $1$. Error bars indicate standard error.}
        \label{fig:4_result_5_cluster_preference}
    \end{figure*}

\subsection{K--means Cluster Analysis} 
\label{sec:4_K_means}

As a result of k-means clustering, we obtained five clusters, A–E, as shown in Fig.~\ref{fig:4_result_5_cluster_preference}. Cluster A had the most participants ($N=377$), and the second largest cluster was cluster B ($N=368$). Clusters C, D, and E had $41$, $29$, and $22$ participants, respectively. 
Each bar in the figure represents each cluster. 
The horizontal axis shows the names of clusters.
The vertical axis shows the normalized preference values. The preference value of a participant for each performance metric was normalized by dividing preference values by the maximum absolute value of the preference and setting the maximum absolute value of the preference to 1. The preference value of each cluster was calculated as the average preference value for each performance metric of participants who were assigned to each cluster. In Fig.~\ref{fig:4_result_5_cluster_preference}, for example, almost all the participants in cluster B prioritized precision. 

% \todo{check the wording of prefer, prioritize, and trivialize}
In the clustering result, there are two big patterns.
% in terms of the preference for fairness metrics, i.e., DI, EO, and CF.
As the first pattern, the participants in the largest two clusters, A and B, did not show any specific tendencies for fairness metrics, i.e., DI, EO, and CF. The participants in A and B did not prioritize nor trivialize fairness metrics. On the other hand, the two clusters had opposite tendencies for the other performance metrics, i.e., accuracy, specificity, sensitivity, and precision. The participants in cluster A prioritized specificity and trivialized precision. In contrast, participants in cluster B prioritized precision and trivialized specificity.

As the second pattern, the participants in clusters C, D, and E had specific tendencies for the fairness metrics. The participants in cluster C prioritized EO and CF and trivialized DI. In contrast, the participants in cluster D trivialized CF and EO, while they did not seem to have a specific tendency in preference for DI. The participants in cluster E trivialized EO, prioritized DI, and did not have a strong tendency toward CF. These three clusters were also different in their preference values for other performance metrics. In cluster C, although the participants' preference values for almost all of the non-fairness performance metrics did not have any strong tendencies, the preference value for sensitivity was relatively higher than other performance metrics. In cluster D, the participants tended to trivialize accuracy, specificity, and sensitivity. In cluster E, the participants prioritized specificity and trivialized sensitivity at the same level as EO.

    \begin{table*}[ht]
        \centering
        \resizebox{\linewidth}{!}{\begin{tabular}{|cc|cc|cc|cc|ccc|}
        \hline
        % header
         & \multicolumn{1}{c|}{} & \multicolumn{2}{c|}{Work status} & \multicolumn{2}{c|}{Education level} & \multicolumn{2}{c|}{Programming years} & \multicolumn{3}{c|}{Income}\\
        % multi header
        \multicolumn{1}{|c}{Clusters}&  & Employee & Employer & \makecell{2 years \\ or less} & \makecell{4 years \\ or more} & \makecell{5 years \\ or less} & \makecell{More than \\ 5 years} & High & & Low\\
        \hline
        % frequency values
        A (N=377) & \makecell{count (\%) \\ lift value} & \makecell{168 (44.6\%) \\ 0.947} & \makecell{209 (55.4\%) \\ 1.047} & \makecell{77 (20.4\%) \\ 1.042} & \makecell{300 (79.6\%) \\ 0.990} & \makecell{246 (65.3\%) \\ 1.011} & \makecell{131 (34.7\%) \\ 0.979} & \makecell{113 (30\%) \\ 0.916} & & \makecell{264 (70\%) \\ 1.041}\\
        B (N=368) & \makecell{count (\%) \\ lift value} & \makecell{179 (48.6\%) \\ 1.033} & \makecell{189 (51.4\%) \\ 0.970} & \makecell{72 (19.6\%) \\ 0.999} & \makecell{296 (80.4\%) \\ 1.000} & \makecell{246 (66.8\%) \\ 1.036} & \makecell{122 (33.2\%) \\ 0.934} & \makecell{122 (33.2\%) \\ 1.013} & & \makecell{246 (66.8\%) \\ 0.994}\\
        C (N=41) & \makecell{count (\%) \\ lift value} & \makecell{18 (43.9\%) \\ 0.933} & \makecell{23 (56.1\%) \\ 1.060} & \makecell{4 (9.8\%) \\ {\cellcolor{lightgray}0.498}} & \makecell{37 (90.2\%) \\ 1.122} & \makecell{18 (43.9\%) \\ {\cellcolor{lightgray}0.680}} & \makecell{23 (56.1\%) \\ {\cellcolor{darkgray}\textcolor{white}{1.581}}} & \makecell{13 (31.7\%) \\ 0.969} & & \makecell{28 (68.3\%) \\ 1.015}\\
        D (N=29) & \makecell{count (\%) \\ lift value} & \makecell{15 (51.7\%) \\ 1.099} & \makecell{14 (48.3\%) \\ 0.912} & \makecell{4 (13.8\%) \\ {\cellcolor{lightgray}0.704}} & \makecell{25 (86.2\%) \\ 1.072} & \makecell{16 (55.2\%) \\ 0.855} & \makecell{13 (44.8\%) \\ {\cellcolor{darkgray}\textcolor{white}{1.263}}} & \makecell{12 (41.4\%) \\ {\cellcolor{darkgray}\textcolor{white}{1.264}}} & & \makecell{17 (58.6\%) \\ 0.872}\\
        E (N=22) & \makecell{count (\%) \\ lift value} & \makecell{14 (63.6\%) \\ {\cellcolor{darkgray}\textcolor{white}{1.352}}} & \makecell{8 (36.4\%) \\ {\cellcolor{lightgray}0.687}} & \makecell{7 (31.8\%) \\ {\cellcolor{darkgray}\textcolor{white}{1.624}}} & \makecell{15 (68.2\%) \\ 0.848} & \makecell{14 (63.6\%) \\ 0.986} & \makecell{8 (36.4\%) \\ 1.025} & \makecell{14 (63.6\%) \\ {\cellcolor{darkgray}\textcolor{white}{1.944}}} & & \makecell{8 (36.4\%) \\ {\cellcolor{lightgray}0.541}}\\
        \hline
        Total (N=837) & count (\%) & 394 (47.1\%) & 443 (52.9\%) & 164 (19.6\%) & 673 (80.4\%) & 540 (64.5\%) & 297 (35.5\%) & 274 (32.7\%) & & 563 (67.3\%)\\
        \hline
        \multicolumn{11}{c}{\vspace{0.1cm}}\\
        \hline
        % header
        & \multicolumn{1}{c|}{} & \multicolumn{2}{c|}{Age} & \multicolumn{2}{c|}{Gender} & \multicolumn{2}{c|}{Race} & \multicolumn{3}{c|}{Political affiliation}\\
        % multi header
        \multicolumn{1}{|c}{Clusters} &  & 18 to 44 & 45 or more & Female & Male & Non-White & White & Democrat & Republican & Other\\
        \hline
        % frequency values
        A (N=377) & \makecell{count (\%) \\ lift value} & \makecell{319 (84.6\%) \\ 0.986} & \makecell{58 (15.4\%) \\ 1.082} & \makecell{120 (31.8\%) \\ 0.958} & \makecell{257 (68.2\%) \\ 1.021} & \makecell{43 (11.4\%) \\ 1.038} & \makecell{334 (88.6\%) \\ 0.995} & \makecell{180 (47.7\%) \\ 0.975} & \makecell{153 (40.6\%) \\ 1.026} & \makecell{44 (11.7\%) \\ 1.018}\\
        B (N=368) & \makecell{count (\%) \\ lift value} & \makecell{323 (87.8\%) \\ 1.023} & \makecell{45 (12.2\%) \\ 0.860} & \makecell{130 (35.3\%) \\ 1.064} & \makecell{238 (64.7\%) \\ 0.968} & \makecell{34 (9.2\%) \\ 0.841} & \makecell{334 (90.8\%) \\ 1.020} & \makecell{184 (50\%) \\ 1.021} & \makecell{143 (38.9\%) \\ 0.983} & \makecell{41 (11.1\%) \\ 0.971}\\
        C (N=41) & \makecell{count (\%) \\ lift value} & \makecell{36 (87.8\%) \\ 1.024} & \makecell{5 (12.2\%) \\ 0.858} & \makecell{14 (34.1\%) \\ 1.028} & \makecell{27 (65.9\%) \\ 0.986} & \makecell{5 (12.2\%) \\ 1.109} & \makecell{36 (87.8\%) \\ 0.986} & \makecell{23 (56.1\%) \\ 1.145} & \makecell{14 (34.1\%) \\ 0.863} & \makecell{4 (9.8\%) \\ 0.851}\\
        D (N=29) & \makecell{count (\%) \\ lift value} & \makecell{24 (82.8\%) \\ 0.965} & \makecell{5 (17.2\%) \\ 1.213} & \makecell{8 (27.6\%) \\ 0.831} & \makecell{21 (72.4\%) \\ 1.084} & \makecell{6 (20.7\%) \\ {\cellcolor{darkgray}\textcolor{white}{1.882}}} & \makecell{23 (79.3\%) \\ 0.891} & \makecell{12 (41.4\%) \\ 0.845} & \makecell{13 (44.8\%) \\ 1.134} & \makecell{4 (13.8\%) \\ 1.203}\\
        E (N=22) & \makecell{count (\%) \\ lift value} & \makecell{16 (72.7\%) \\ 0.848} & \makecell{6 (27.3\%) \\ {\cellcolor{darkgray}\textcolor{white}{1.918}}} & \makecell{6 (27.3\%) \\ 0.821} & \makecell{16 (72.7\%) \\ 1.089} & \makecell{4 (18.2\%) \\ {\cellcolor{darkgray}\textcolor{white}{1.654}}} & \makecell{18 (81.8\%) \\ 0.919} & \makecell{11 (50\%) \\ 1.021} & \makecell{8 (36.4\%) \\ 0.920} & \makecell{3 (13.6\%) \\ 1.189}\\
        \hline
        Total (N=837) & count (\%) & 718 (85.8\%) & 119 (14.2\%) & 278 (33.2\%) & 559 (66.8\%) & 92 (11\%) & 745 (89\%) & 410 (49\%) & 331 (39.5\%) & 96 (11.5\%)\\
        \hline
        \end{tabular}}
        \caption{The table presents the results of association analysis, using lift values to quantify the strength of the association between demographic attributes and each cluster. A lift value greater than 1 indicates a positive association between the attribute and the cluster, while a value smaller than 1 indicates a negative association. Cells with lift values below 0.750 are highlighted in light gray, and cells with lift values greater than 1.250 are highlighted in dark gray with white text for emphasis.}
        \label{table:4_result_frequency_table}
    \end{table*}

\subsection{Association Analysis by Lift Value} \label{ssec:4_Association_Analysis}

\takuya{15. Need additional analysis: DONE}
Our analysis explored the association between demographic attributes and the clusters. Table~\ref{table:4_result_frequency_table} shows the numbers, percentages, and lift values of participants who had each attribute value in each cluster.
We initially focus on attributes with lift values greater than 1.250 because these indicate a strong association between the demographic attributes and the clusters. Then, we focus on attributes with lift values below 0.750 to identify negative associations with the clusters.
Note that except for the work status, there were imbalances in the distribution of participants across attributes. Political affiliation and gender did not show a positive or negative association with any of the clusters.

Although most attributes in clusters A and B had lift values close to 1, indicating a weak association, clusters C, D, and E showed various tendencies in lift values.
Cluster C was associated with more programming years (lift value of 1.581). On the negative side, cluster C had a low lift value of 0.498 for participants with ``2 years or less'' of education and 0.680 for fewer programming years, suggesting a negative association with these attributes.
Cluster D had an association with more programming years (1.263) and high income (1.264).
The cluster also had a strong association with a non-white racial background (1.882). On the other hand, it had a low lift value of 0.704 for participants with "2 years or less" of education, suggesting a negative association.
Cluster E was diverse in its associations. It had positive lift values for being an employee in work status (1.352). Also, low educational level (1.624), high income (1.944), older age (1.918), and non-white (1.654) are related to the cluster. On the negative side, cluster E showed a lift value of 0.541 for participants with low income and 0.687 for those who are employers.

\section{Discussion}

% \todo{add what should be considered when we try to execute a multi-stakeholder evaluation of ML models based on this study > yuri}

% \todo{modify}
We discuss our results from the perspectives of our research questions. We then discuss the implications and limitations of our study.

\subsection{Preferences for Metrics} 
% \subsection{Summary of the Findings Related to Research Question 1} 
\label{ssec:5_summary}

% 4.1 クラスターと選好について -> RQ1
% AとBが対照的
% CとEが対照的
To address RQ1, we clustered participants' preferences with k-means clustering (Fig.~\ref{fig:4_result_5_cluster_preference}). The graphs of clusters A and B were seemingly in contrast. The participants in cluster A trivialized precision and prioritized specificity and accuracy, while those in cluster B showed the opposite tendency. While clusters C and E did not indicate a clear difference in pattern compared to clusters A and B, clusters C and E seemed to be in contrast to the preferences for sensitivity, DI, and EO. 
%  Cluster C prioritized sensitivity, EO, and CF and gave less priority to specificity and DI, while cluster E prioritized specificity and DI and trivialized sensitivity and EO. 
Cluster D, however, did not show a contrasting pattern from the others, yet it trivialized CF the most among all clusters.

Some of these tendencies are based on the relationship among the seven performance metrics. The relationships between sensitivity (or recall) and specificity are in a trade-off relation. If sensitivity increases, specificity decreases, and vice versa. This suggests that there are similarities in the tendencies of clusters A and E, i.e., relatively high preference values for specificity and relatively low preference values for sensitivity, and clusters B, and C, i.e., the opposite tendency to clusters A and E for the preferences for specificity and sensitivity. Note that EO is a measure of the demographic parity of sensitivity. Therefore, a participant can prioritize sensitivity and EO simultaneously. This suggests that it is natural that, in all clusters, the preferences for sensitivity and EO are similar. Therefore, it was difficult to determine whether the participants prioritized or trivialized sensitivity or EO on the basis of only this analysis. 

% On the other hand, CF, in our definition, does not have a clear relationship with any other metrics.

% As for recall (or sensitivity) and precision, their scores would become equal when the proportion of false negatives and false positives is equal. 
% Recall increases and precision decreases if false negatives decrease from that equal point, and precision increases and recall decreases if false positives decrease.

% Regarding relationships between the seven metrics, they can have trade-offs.

% 結論
Overall, the participants' preferences for metrics were divided into recognizable patterns. The main pattern was whether the participants had specific preferences for the fairness metrics (clusters, C, D, and E), or not (clusters A and B).  The majority of the participants were in the clusters A and B. In the clusters in which the participants had specific preferences for fairness metrics, different clusters had different preferences for each fairness metric. Clusters C and D had opposite tendencies, especially toward CF. Additionally, despite DI being one of the most well-known fairness metrics~\cite{barocas2016big}, only the participants in cluster E tended to prioritize DI. It is also notable that the participants in the clusters without any specific preference for fairness had opposite tendencies to precision and specificity.
% \todo{need more explanation?}

% Therefore, regarding RQ1, it was possible to quantify the preferences for metrics of each participant using the questionnaire-based method.

\subsection{Demographic Attributes}
% \subsection{Summary of the Findings Related to Research Question 2}

% \takuya{15. Need additional analysis, add descriptions about lift value: DONE}

To address RQ2, we compared the differences in demographic attributes between clusters using association analysis. We analyzed the relationship between demographic attributes and clusters, as detailed in Table~\ref{table:4_result_frequency_table}.
Clusters C, D, and E are particularly noteworthy because they showed distinct patterns in both fairness metrics and demographic attributes.

Our analysis shows that educational attributes influence membership in cluster C. Specifically, having a lower education level reduces the likelihood of being in cluster C by half (lift value of 0.498). Similarly, having fewer programming years decreases the likelihood by over 25\% (lift value of 0.680). On the other side, more programming years increase the likelihood of being in cluster C by more than 1.5 times (lift value of 1.581). This suggests that cluster C is associated with educational and programming experience. Most participants in this cluster prioritize CF and are experienced programmers. This trend suggests that individuals who believe that job expertise should determine hiring are actually skilled. 
Additionally, this cluster also prioritizes sensitivity, along with the EO, which aims to reduce false negatives. This shows the idea of making fewer errors to ensure that qualified candidates get hired.
The trivialization of DI by this cluster suggests that the cluster does not place emphasis on racial disparities in the likelihood of being predicted to be hired.

Regarding cluster D, having lower education levels reduces the likelihood of belonging to cluster D by over 25\% (lift value of 0.704). On the other hand, being a non-white individual increases the likelihood of belonging in this cluster by more than 1.8 times (lift value of 1.882). Participants in cluster D tend to trivialize CF and EO, and they did not show a clear preference for DI. This suggests that people in this cluster may not believe that higher job expertise should be the only reason for getting hired. Similarly, the trivialization of EO suggests they may not be concerned with reducing false negatives. They also trivialize specificity. This suggests they are not focused on minimizing false positives. Since they do not clearly prefer any other metrics, there is room for interpretation. For instance, they might believe that hiring decisions should not be based solely on attributes like race and expertise.

Participants in cluster E prioritize DI, trivialize EO, and show a relatively small trend for CF while also prioritizing specificity. The preference for DI suggests they are concerned about racial disparities in the likelihood of being hired. The trivialization of EO indicates they are not aiming to reduce false positives. On the other hand, prioritizing specificity implies they want to minimize the risk of hiring unsuitable candidates. Demographic attributes positively associated with this cluster include being an employee (lift value of 1.352), having a lower level of education (1.624), higher income (1.944), older age (1.918), and being non-white (1.654). Being low-income halves the likelihood of belonging to this cluster (0.541), and being an employer reduces it to 0.687. The prioritization of specificity suggests they may be seeking a beneficial state for employers by reducing the risk of false positives. The high lift values for higher income and older age indicate that they might hold higher job responsibilities.
However, these lift values should be interpreted cautiously, especially considering the small size of cluster E, which has only 22 participants. 
% For example, while higher income and older age are not empirically contradictory, finding a reason for the association between being an employee, having a lower level of education, and being non-white is challenging.

\subsection{Minority Group}

% - 多数派と少数派の違い
%   - 多数派は精度、中でもprecisionについての意見が真っ二つ。
%   - 少数派は公平性とか気にしがち
% - 少数派の意見を取るときに誰に注意すべきか
%   - リフト値の分析から分かることを書く。
In this study, we emphasize examining preferences within both majority and minority groups because the focus of this study is to give due consideration to minority opinions. 
Utilitarian methods, such as averaging preferences, may be sufficient for calculating the total utility of all stakeholders. However, in the context of fairness, it is critical to focus on demographic minorities and explore ways to secure their interests. Utilitarian approaches to aggregating preferences could inadvertently marginalize minority groups if the majority does not consider fairness metrics.

We observed clear differences among majority and minority groups in their preferences for metrics and tendencies in demographic attributes. Clusters A and B, the majority clusters, consist of 89\% of the participants, and they show interest only in performance metrics. Cluster A prioritizes precision and trivializes specificity, while cluster B does the opposite. Both clusters aim to reduce false positives but differ in whether they want to increase true positives or true negatives. On the other hand, minority groups, represented by clusters C, D, and E, are generally concerned with fairness. Related to the tendencies in demographic attributes, cluster C correlates with education level and programming years, possibly explaining a preference for individual fairness CF. Additionally, cluster D is associated with education level, programming years, and race, and cluster E is associated with work status, educational level, income, age, and race. These tendencies in demographic attributes clarify that it is effective to pay attention to demographic attributes when characterizing minority clusters. On the other hand, it should also be noted that the smaller size of minority groups poses challenges for detailed quantitative analysis.

% The importance of focusing on minority opinions is supported by the analysis of lift values. \yuri{$\leftarrow$what this means?$\rightarrow$}\textbf{This analysis reveals that demographic biases exist in the preferences of minority groups, emphasizing the need to pay attention to stakeholders with specific demographic attributes when considering minority opinions. }

\subsection{Implications towards Multi-Stakeholder ML Evaluation} \label{ssec:5_implications} 
\textbf{Checking attributes that represent minorities in ML model evaluation:}
When evaluating ML models with multiple stakeholders, it should be checked if there are specific attributes that represent the minority group in the sense that their evaluation of the ML model differs from others.
This is because the results of this study indicate that stakeholder preferences can be clustered and that general demographic attributes and domain-specific attributes such as work status are associated with the tendencies in minority clusters.
% Moreover, only in the minority clusters are there clear tendencies in the demographic and domain-specific attributes. 

\textbf{Explain to the group that gives different evaluations to the ML model to make them aware of the different values: }
When it is necessary to balance the trade-offs between each cluster's preferences, different explanations should be provided in a way that balances preferences.
Our results indicate that many people are generally more interested in performance than fairness metrics.
To ensure group or individual fairness, we may need to explain the importance of improving fairness metrics to such people. There also may be different attributes among those who have specific preferences for fairness metrics.

% it should be noted that when making decisions affecting multiple stakeholders through ML, it is necessary to balance the trade-offs between each cluster's preferences and explain the decisions in a way that balances them.  

\textbf{Recruit a diverse group of people when evaluating ML models: }
The diversity of the people who evaluate ML models relates directly to the output of the evaluation. And to ensure the diversity of those who evaluate models leads to a fair model evaluation process.
In the domain of job matching, it is important to recruit diverse evaluators regarding programming years or income, as revealed in our experiment.
For example, the results of this study indicate that those who prioritize applicants with higher skills tend to be those with more programming years, while those who prioritize group fairness tend to be those with higher incomes. The former, in particular, aligns with the commonly observed tendency to favor people with attributes similar to one's own. 
Based on this results, because it is difficult for people to ignore their own interests, diverse people should participate in the model evaluation.

% Future research should develop methods that extract more explanatory information while ensuring as much simplicity of the task and demographic questions as possible to avoid cognitive overload.

% Finally, this study lacks qualitative analysis. 
% Future research can extend the interpretation of the reason behind participants' preferences by systematically incorporating both quantitative and qualitative analyses.

\subsection{Limitations and Future Work} \label{ssec:5_limitation}

Although our results indicate associations between clusters and demographic attributes, they do not clarify a causal relationship. We need to conduct a qualitative analysis to understand the reasons behind the participants' choices. This will help us interpret the reasons for the patterns of metric prioritization that we found. 
% \takuya{\#17 O6. "94.8\% of participants have experience(s) of developing one or more machine-learning models" -- this proportion seems quite high compared to an average stakeholder who might interact with a job-matching ML model. What are the implications of this bias in study participants? }
In our study, we found that 94.8\% of participants had experience in developing one or more ML models. However, in real-world systems, it is likely that a larger proportion of individuals have limited knowledge of ML. This could potentially impact the results. 
In such scenarios, therefore, the preference patterns may not be as easily discernible compared with the findings outlined in this paper because of the unfamiliarity with ML, which may lead to difficulties in interpreting the scenario and setting. 
% This is primarily because individuals who are unfamiliar with ML might struggle to identify and interpret the underlying patterns within the data, resulting in less pronounced patterns in their preferences across different tasks. 

Next, this study was limited to collecting only basic demographic information. 
% while we tried to explain stakeholders' preferences by using their demographic attributes as proxies.
To identify the crucial demographic attributes for clustering participants, we need to carefully consider the appropriate attributes that ensure diversity among the participants.
Similarly, the small sample size in minority clusters poses challenges for cross-validating these attributes, indicating a need for more robust methods to discover inter-attribute relationships.
% Particularly, the small sample size of cluster E, which consists of only 22 participants. This makes it difficult to draw definitive conclusions about the associations between being an employee, having a lower level of education, and having a non-white racial background.
% To address this, 
Future research should aim to include a larger and more diverse sample to validate these findings.

Finally, when considering the generalizability of the procedure and study results, one limitation is that the current procedure focuses only on two features (race and job expertise) with two values each. Real-world scenarios often involve multiple features with varying values. 
Applying our method to an ML model used for the classification task for three or more classes would require transforming the classification results into binary or significantly increasing the number of preference-extraction tasks to determine accurate individual preferences. For this purpose, it is crucial to consider how to transform the classification results into binary and how to address the cognitive load imposed by these modifications. 

\section{Conclusion}
Toward evaluating ML systems, including the values of multiple stakeholders, we proposed a crowdsourcing method to investigate how the stakeholders' preferences for the metrics are clustered and with what attributes stakeholders in each cluster have.
We conducted an experiment in the context of job matching and obtained empirical results from 837 participants, clustered them with their preferences, and analyzed the tendencies in the participants' demographic attributes in each cluster. We found that there are clusters of the majority of people who have clear tendencies on the performance metrics, such as specificity or precision, and clusters of minority people who have characteristics of preference for the fairness metrics.
To make multi-stakeholder evaluation more practical, it is crucial to develop the model of stakeholders based on their preferences for the metrics and the stakeholder groups based on their demographic attributes.

\section{Acknowledgements}
This version of the contribution has been accepted for publication, after peer review (when applicable) but is not the Version of Record and does not reflect post-acceptance improvements, or any corrections. 
The Version of Record is available online at: http://dx.doi.org/[insert DOI]. 
Use of this Accepted Version is subject to the publisher’s Accepted Manuscript terms of use https://www.springernature.com/gp/open-research/policies/accepted-manuscript-terms.

\bibliographystyle{splncs04}
\bibliography{bibliography} % REFERS TO "bibliography.bib"

\end{document}